# An Advanced Conceptual Diagnostic Healthcare Framework for Diabetes and Cardiovascular Disorders

M. Sharma,[1]*,G. Singh[2] and R. Singh[2]

[1]DAV University, Jalandhar, India.
[2]Guru Nanak Dev University, Amritsar.

## Abstract

The data mining along with emerging computing techniques have astonishingly influenced the healthcare industry. Researchers have used different Data Mining and Internet of Things (IoT) for enrooting a programmed solution for diabetes and heart patients. However, still, more advanced and united solution is needed that can offer a therapeutic opinion to individual diabetic and cardio patients. Therefore, here, a smart data mining and IoT (SMDIoT) based advanced healthcare system for proficient diabetes and cardiovascular diseases have been proposed. The hybridization of data mining and IoT with other emerging computing techniques is supposed to give an effective and economical solution to diabetes and cardio patients. SMDIoT hybridized the ideas of data mining, Internet of Things, chatbots, contextual entity search (CES), bio-sensors, semantic analysis and granular computing (GC). The bio-sensors of the proposed system assist in getting the current and precise status of the concerned patients so that in case of an emergency, the needful medical assistance can be provided. The novelty lies in the hybrid framework and the adequate support of chatbots, granular computing, context entity search and semantic analysis. The practical implementation of this system is very challenging and costly. However, it appears to be more operative and economical solution for diabetes and cardio patients.

*Corresponding author. Email:manik_sharma25@yahoo.com

## 1. Introduction

Diabetes is a non-contagious chronic human disorder. Early prognosis of this disorder can expose the grievous complications and help to save human life. The shattering increase in obesity and torpid lifestyle has made diabetes a universal epidemic. The primary reason behind this disease is the overdose of glucose in the blood. Human body changes most of the food into glucose, and it is the primary source of energy in the body. A person becomes diabetic when his/her body is not able to transform glucose into energy. Due to excessive glucose in the blood, BSL (Blood Sugar Level) increases that ultimately leads to diabetes [1][2]. Numbers of individuals are getting affected by diabetes. In the year 2000, India topped the world with 31.7 million diabetic people followed by China with second and United States at third place. It is predicted that by the year 2030 diabetes mellitus may affect up to 79.40 million people in India. According to the WHO (World Health Organization), it was estimated that 3.4 million deaths are caused due to high blood sugar [3].

Cardiovascular disease is a major fatal human disorder. The major cardio disorders can affects the working of the coronary artery, valves function, heartbeat, heart chamber and other structural component of the heart. The common indicators for these heart disorders are chest pain, indigestion, vomiting, sweating, nausea, high fatigue, breathlessness and skin rashes. High blood pressure, smoking, drinking, hereditary, age, overweight are the common reasons that may lead to these heart disorders. It has been observed that Turkmenistan and Ukraine are on the top when cardiovascular causalities are of concern. However, in regard to the cardiovascular disorder casualties, Vietnam, Japan, France and South Korea are on the safest side. There are different ways to diagnose this deadly human disorder. In the last two decades, several DM techniques have been employed for early diagnosis of this disease. From Figure 1, is observed that Indians are mostly affected by cardio disorders. The rate of causalities due to cardio disorder is 1383.6 per million, which is obviously very high as compared to other human disorders[4][5].

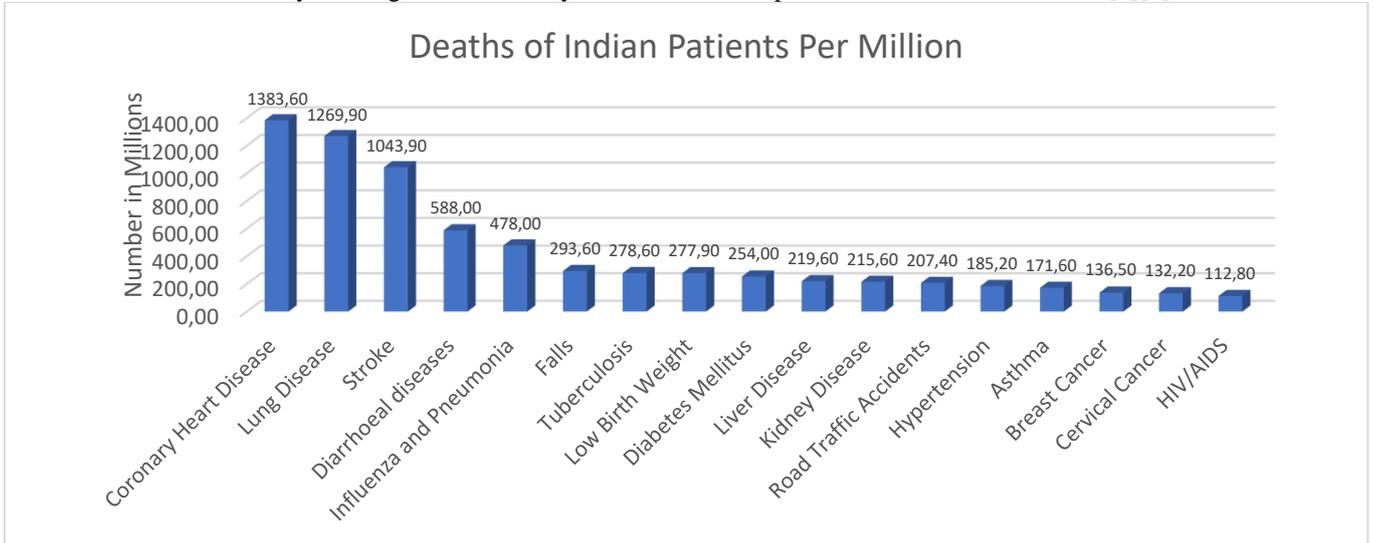

**Figure 1 : Different Diseases and the Death Rate**

Numbers of data mining based diagnostic solutions are available for diagnosis of different human disorders like cancer, diabetes, heart and neurological disorders. Thongkam J et al. have proposed a C-SVCF (C-Support Vector Classification Filter) that assists in finding and removing the misclassified instances of breast cancer data set. Authors found that the performance of C-SVCF is outstanding as compared to the ensemble-based filters [6]. Similarly, Bailing Zhang et al. have proposed a Bayesian model-based solution for mining of Cerebral palsy (CP) which is considered as a non-progressive neuro-developmental condition. Authors found better classification problem with the Bayesian-based model as compared to SVM and other neural network models [7]. Saravananathan and T. Velmurugan have employed different classification techniques such as J48, CART, SVMs, and kNN to mine the data of diabetic patients. Difference performance metrics were computed. Authors found outstanding classification rate with J48 as compared to other classification techniques [8].

Several queries have been fired to Google Scholar to examine the publication trend, and it is observed that a significant number of articles related to diabetes and human cardiovascular disorders have been indexed. It has been noted that a maximum number of researchers have employed different data mining techniques to diagnose cardio disorders. Figure 2 depicts the diseases and the number of an article indexed for mining of those disorders using various data mining techniques.

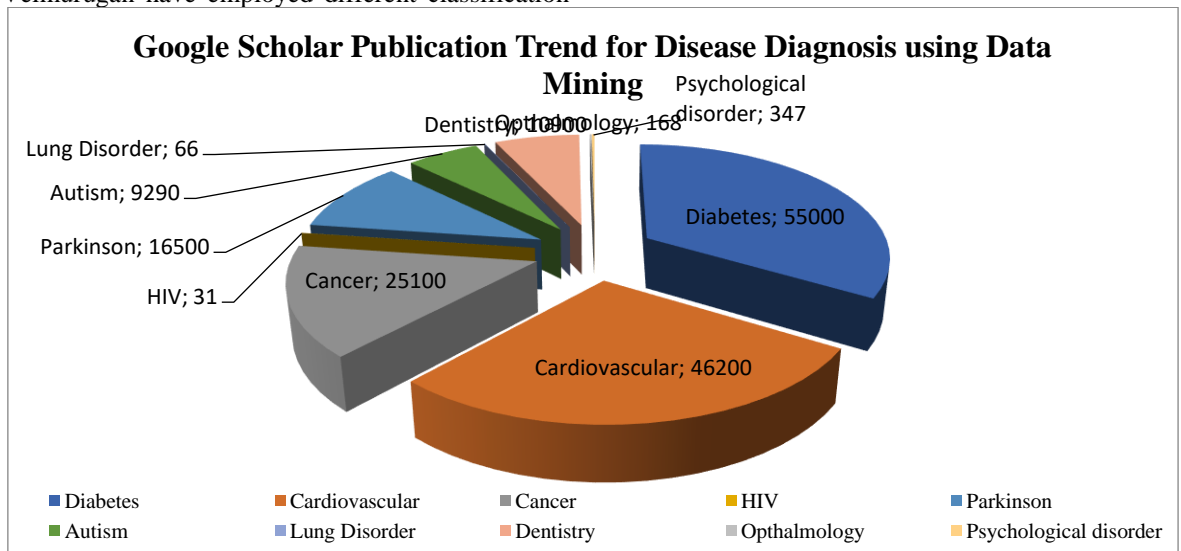

**Figure 2: Google Scholar Publication Trend for Disease Diagnosis using Data**

Here, a smart conceptual framework has been proposed for the diagnosis of diabetes and human cardiovascular disorders. The smartness of the system is based on the ability to collect, process, and transform healthcare data into valuable information so that the periodic data of the patients can be regularly monitored. However, the implementation of these systems is a challenging task [9].

The proposed hybrid healthcare system has been designed using an amalgamation of distinct emerging computing techniques like machine learning, Internet of Things, Chabot's, context entity searches and granular computing. In this system, the bio-sensors and IoT are intended to collect and monitor the regular and periodic data of diabetic and heart patients. Sentiment analysis techniques are used to mine different information related to these critical human disorders. Data mining and machine learning techniques are used to classify diabetic and cardio patients from the healthy people. Furthermore, granular computing assists to focus on the critical area of disorder diagnosis. The reports will be represented using different data visualization techniques. In spite of these factors, telehealth and smart tools can be used to avail 24x7x365 online facility where a person or machine will communicate with the patients and try to remove the issues of the patients.

## 2. Literature Review

This section will highlight the works of some of the key researchers who designed different diagnostic healthcare framework for diagnosis of diabetes and cardio patients.

## 2.1 IT-based healthcare solution for diabetes

Several IT-based diabetes diagnostic frameworks have been designed by using different computation techniques. The statistical techniques based diabetes diagnostic model has shown better predictive results using decision tree [10]. C4.5 seems to be preferred choice as the rate of accuracy accomplished using C4.5 was enriched than other classification techniques [11][12]. For gestational diabetes diagnosis, random forest outperforms other classification procedures. The good rate of accuracy (93.5%) has been attained using random forest-based gestational diagnostic system [13]. Smartphone's are also playing valuable role in diabetes care. It has witnessed that more than 70% health applications are related to diabetes. These applications are supporting to reduce the affected rate of diabetes patients [14]. Authors used microcontroller that scans the patient's data using sensor and then Wi-Fi module send the pre-processed data to the doctors for recommendation of the patients. The system is competent to confine heart rate and blood pressure [15]. An IoT based diabetes care system uses the services of glucose general packet radio services and blood-glucose monitor to extract data from diabetic patients. The data collected from the patients is uploaded to cloud server. Healthcare professional are able to access the data uploaded over cloud server. Whenever an abnormal data is detected the system sends the notification to both patients as well as doctors [16]. One of the most important parameters for IoT based healthcare system is the user interface. The system should support sound medical expertise as compared to other features. The online support should be from healthcare professional rather than call centre executives. Additionally system should be recommended by medical regulatory bodies [17]. The use of HTML5 will further improve the performance of m-healthcare applications. By using HTML5, one is able to develop a single application that can be effectively run over different mobile and desktop versions without any modification. In addition, the HTML5 based m-health application have smoother interface [18]. The use of social media in sensing and disseminating healthcare information is exponentially increasing. Semantic analysis and ontologies are used to extract meaningful information from massive volume of online posts. Ontology builds a relationship between patients, symptoms, disorder, prognosis and treatment of disease. Additionally, people are using Facebook, Twitter, Whatsapp and other social applications as an admired source for healthcare. In Asia, more than 30% users are seeking medical assistant from social media or healthcare web applications [19].

## 2.2 IT-based healthcare solution for cardiovascular disorders

Ilayaraja M., Meyyappan has developed a method to foresee the peril of heart disease using frequent itemset. Author has taken 19 different symptoms which are the reasons behind heart disease. Authors evaluated the consequences of proposed techniques with apriori algorithm, IMSIA algorithm, semi-apriori algorithm and claimed that their algorithm works well as compared to other existing techniques [20]. Purushottam et al. (2016) have designed original, pruned, non-duplicates, classified and polish rules for diagnosis of cardio disorders. Authors compared the accuracy of their efficient diagnostic system with other mining techniques like SVM, C4.5, MLP, RBF and neural network and found that the predictive accuracy obtained using proposed system is better than other methods [21]. B. Venkatalakshmi, M.V Shivsankar stated untimely diagnose of a heart problem might lead to illness or even death in some cases. Authors diagnosed the cardio disorders using two distinct data mining techniques viz. naive Bayes and decision tree. Authors found that naive Bayes gives optimal results as compared to the decision tree. The accuracy of Naïve Bayes is 85.03% [22]. Taneja has developed a prophetic model to mine Transthoracic Echocardiography Report dataset. Author has collected data from one of the reputed hospital of Chandigarh.

Three different supervised machine learning techniques viz. J48 classifier, multilayer perception and Naive Bayes have been used. The accuracy of classification obtained was 95.56[23].

Dangare and Apte stated that data mining techniques are beneficial in predicting the heart disease based upon the history of the patients. Authors used and compared the performance of different data mining techniques in predicting the status of heart disease. Authors found that neural network gives 100% accuracy in determining the heart disease status [24]. Sudha et al. stated that stroke is a life-threatening disease and is one of the major causes behind the deaths. Authors tried to analyze the data of stroke patients. They used different data mining techniques viz. neural network, Naïve Bayes and decision tree. The author got better results with the neural network. Authors proposed a model that predicts the nature of data based upon the feature subset selection. The accuracy of the proposed framework is 91% [25]. Mamta Sharma et al. (2017) have compared different data mining techniques used to predict heart diseases. Authors claimed that the predictive rate of the neural network in diagnosing heart related human disorders is better than the Naïve Bayes and decision tree [26]. Ebenezer Obaloluwa Olaniyi et al., (2015) have experimented with one of the UCI heart disease related datasets. Authors compared the results of SVM and feed-forward multilayer perceptron model. The rate of accuracy achieved using SVM is 2.7% better than the feed-forward multilayer perceptron model [27]. V. Krishaiah et al. have developed a fuzzy KNN based predictive system for heart diseases. Authors used fuzzy approach to remove the uncertainty and to improve the accuracy level of prediction related to heart patients[28]. Bashir et al. have proposed an innovative ensemble classifier based upon five different techniques viz. naïve bayes, decision tree induction based on Gini index, information gain, memory based learner and SVM. Authors approach (MV5) found to be an improve one. The rates of forecasting accuracy, sensitivity and specificity attained using MV5 are 88.52%, 86.96% and 90.83% respectively [29]. Nishara et al. employed a hybridization of C4.5, K-means and maximum frequent itemset to for heart disease diagnosis. They found their hybridized model witnessed more potential as compared to other techniques. The rate of recall, precision and accuracy achieved using their hybrid framework is 0.89, 0.82 and 89% respectively [30].

Maryem Neyja et al. have proposed an IoT based system for cardio patients. The system can get the ECG of the remote patients and based on requirements; it can send the signal to nearest hospitals as well as the concerned patients so that his/her life can be saved [31]. Moeen Hassanalieragh et al. have proposed an IoT-based healthcare model that can collect data from remote patients using different data acquisition and data transformation facilities. To cater to the long-term storage requirements, authors used cloud storage and cloud processing. Authors have also highlighted some of the challenges that need to face for the implementation of their proposed model [32].

## 3. Methodology

Data mining is a multistep approach that is used to for automatic extraction of veiled patterns. Firstly, data is collected from assorted sources and is transformed into a consistent format. In the second phase, the data mining procedures are applied to excavate some meaningful information. The third phase analyzes the processed data and represents it in a standardized format. Finally, the upshots of data mining progression are used in assessment making process. In general, data mining is used to test a hypothesis or to discover some new or hidden patterns. The inspiration was first to organize a statement that has to be tested against a meticulous set of data and condition. Decision tree induction (DTI), Naïve Bayes, rule-based classification, support vector machine and genetic algorithm are some of the major DM techniques. One can also predict the future by carefully analyzing the past and current state of the dataset. The complete working of data mining techniques is based upon two sets of data, i.e. Training and Testing. Data mining is a complicated process as one has to train the system regarding the characteristics or the features that have to be extracted [5]. As of now, some IT-based diagnostic and DSS have been designed for cardiovascular and diabetic patients. DM witnessed a great practical value for developing automated healthcare system. DM devises an intelligently automated procedure that assists in classification, clustering and prediction of disease. It has an outstanding role in early diagnosis and effective treatment for diabetes patients. The different techniques like classification (Naive Bayes, decision tree, genetic algorithm (GA), support vector machine) and clustering (K-mean, GA) procedures have effectively used for pre-diagnosis of diabetes[33].

The advent of machine learning and IoT (Internet of Things) is and will bring world-shattering transformations in the diabetes healthcare system. IoT is a rising technology that connects different entities like computers, laptop, buildings, ambulances, sensors, cars, refrigerators. These are unified to collect, process and disseminate data. IoT is leading to well-built inroads for the smart diabetes healthcare systems. The objective of IoT enabled healthcare system is to offer a comprehensive, intelligent, economical, effective, omnipresent and always connected framework [34].

Internet of Thing (IoT) is an emerging technology that connects several physical devices, sensors, software, embedded devices to collect process and disseminate data. It is a network of computers, laptops, smartphones, people, TV, wearable devices, sensors, medical instruments, control devices, display devices etc. [35][36]. The primary objective of IoT is to provide a platform where a different object from the physical or cyber world can exchange data or information with one another. There are three major pillars of IoT framework are physical or cyber world objects, communication network and analytics tools [37]. IoT is used in different domain like smart cities, Agriculture, energy engagement, healthcare, transportation, connected cars, entertainment etc.

Sentiment analysis, granular computing, chatbots and contextual entity search are some emerging smart tools. From the last few decades, the web opinion mining is growing in leaps and bounds. Users are posting their views or opinions more frequently on the daily basis. It is not feasible for a human being to read all massive online reviews about some important event or object posted by thousands or millions of users. Semantic analysis assist in mining polarity based convincing information from massive data volume of online posts [38]. Granular computing (GC) is a sub-domain of granular mathematics. It is a natural way to solve a complex problem. The roots of GC lie in human thinking. GC states that one has to work directly on granules (fragments) rather than the complete problem. It operates in two modes. In GC, one may start with fuzzy side and move downward or start with the crisp side of the problem and move upwards [39]. Chatbot is an intellectual simulated chatting program where a machine interacts with the user. The machine is trained enough to interact effectively on a particular domain. Chatbots are covering a wide spectrum of application right from entertainment to edutainment and healthcare. Contextual entity search is an automated intellectual search mechanism that automatically opens relevant links and information, whenever the text is selected or highlighted. In addition, it provides the facility of scope searching and query time content mining. Performance, flexibility and intelligence are three major features of context entity search.

## 4. Design of Advanced Conceptual Healthcare Framework (SMDIoT) for Diabetes and Cardio Patients

Researchers have used different data mining and machine learning techniques to provide an automated solution for diabetes and cardiovascular diseases. Recently, IoT has also been used as a major component for these types of healthcare systems. It is observed that very little effort has been put into practice to design a smart, complete and remote monitoring healthcare system for both diabetes and human cardiovascular disorders. However, a variety of IT-based solutions for these disorders are available. A smart healthcare system is needed to provide therapeutic suggestions to the individual victims of diabetes and cardio disorders.

For designing a novel conceptual framework, the publication trends of different healthcare diagnostic solutions have been examined. Table x and y represents the query and a corresponding number of articles indexed in the Google Scholar.

**Table 1: Publication Trend of Cardiovascular disorders related articles**

| cardiovascular disorder | 2110000 |
| --- | --- |
| cardiovascular disorder diagnosis | 43 |
| heart disease diagnosis | 7170 |
| "heart disease diagnosis" + "data mining" | 1480 |
| "heart disease diagnosis" + "Internet of Things." | 93 |
| "heart disease diagnosis" + "machine learning" | 1480 |
| "heart disease diagnosis" + "granular computing" | 45 |
| "heart disease diagnosis" + "sentiment analysis" | 32 |
| "heart disease diagnosis" + "chatbot" | 1 |
| "heart disease diagnosis" + "context entity search" | 0 |
| "heart disease diagnosis" + "context entity search"+"granular computing" | 0 |
| "heart disease diagnosis" + "context entity search"+"machine learning" | 0 |
| "heart disease diagnosis" + "context entity search"+"machine learning" + "internet of things" | 0 |
| "cardiovascular disorder diagnosis" + "data mining" | 3 |
| "cardiovascular disorder diagnosis" + "Internet of Things" | 0 |
| "cardiovascular disorder diagnosis" + "machine learning" | 4 |
| "cardiovascular disorder diagnosis" + "granular computing" | 0 |
| "cardiovascular disorder diagnosis" + "sentiment analysis" | 0 |
| "cardiovascular disorder diagnosis" + "chatbot" | 0 |
| "cardiovascular disorder diagnosis" + "context entity search" | |
| "cardiovascular disorder diagnosis" + "context entity search"+"granular computing" | 0 |
| "cardiovascular disorder diagnosis" + "context entity search"+"machine learning" | 0 |
| "cardiovascular disorder diagnosis" + "context entity search"+"machine learning" + "internet of things" | 0 |

**Table 2: Publication Trend of Diabetes-related articles**

| | |
|---|---|
| diabetes | 2750000 |
| diabetes diagnosis | 30500 |
| diabetes diagnosis + "data mining" | 1180 |
| diabetes diagnosis + "Internet of Things." | 85 |
| diabetes diagnosis + "machine learning" | 1720 |
| diabetes diagnosis + "granular computing" | 36 |
| diabetes diagnosis + "sentiment analysis" | 44 |
| diabetes diagnosis + "chatbot" | 1 |
| diabetes diagnosis + "context entity search" | 0 |
| diabetes diagnosis + "context entity search"+"granular computing" | 0 |
| diabetes diagnosis + "context entity search"+"machine learning" | 0 |
| diabetes diagnosis + "context entity search"+"machine learning" + "internet of things" | 0 |

From the Table 1 and Table 2, it is observed that no significant work has been done using the hybridization of data mining, IoT and other smart tools like chatbot, context entity search, granular computing etc. Therefore, here, a smart data mining and IoT (SMDIoT) based advanced healthcare system for proficient diabetes and cardiovascular diseases have been proposed. The hybridization of data mining and IoT techniques are supposed to give an effective and economical solution to diabetes and cardio patients. The various components of SMDIoT are Patients, healthcare professionals, bio-sensors, chatbot, context entity search, semantic analysis, insulin and cardiovascular sensors, healthcare machines, data mining procedures, granular computing, social network sites, data analysis and visualization tools. The novelty lies in bringing different techniques in a common framework. Moreover, the use of chatbots, semantic analysis and granular computing make this framework more effectual and inventive.

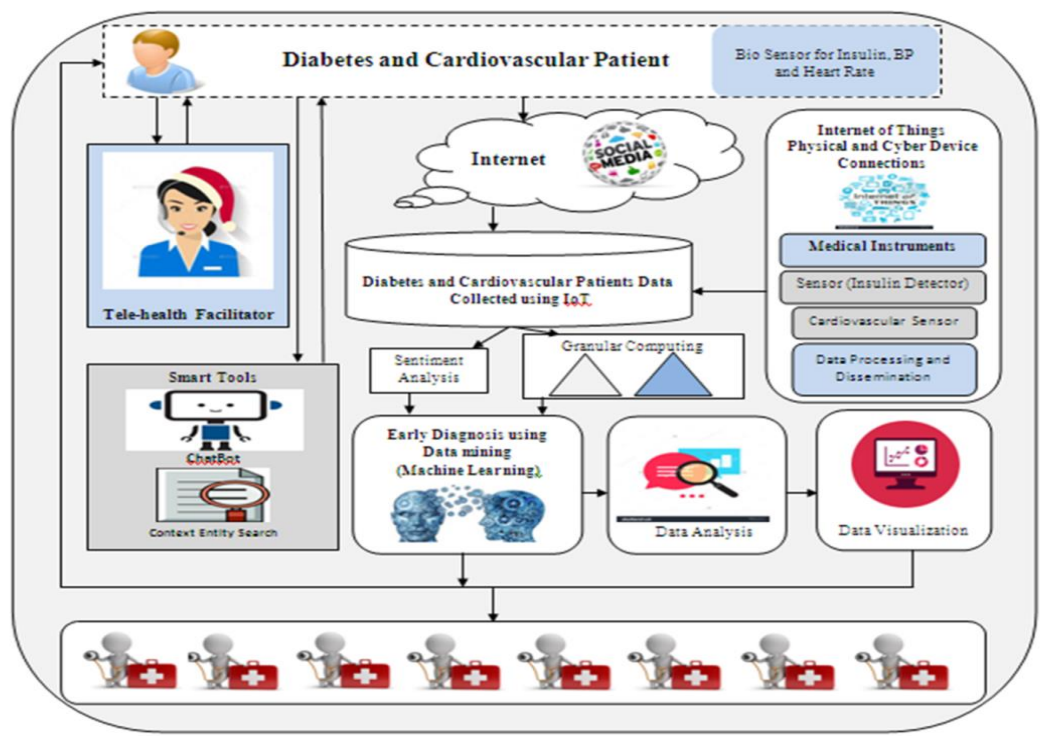

**Figure 3: Data Mining and IoT based Smart and Advanced Healthcare System (SMDIoT)**

The proposed model (SDMIoT) is a smart healthcare solution for complete diabetes care. With SMDIoT, one is able to collect, store, analyze and disseminate complete data related to the health status of diabetes and heart patients. The system is able to provide the smart facility context entity search with which one can quickly and precisely explore substantial online information. Additionally, chatbot and tele-facilities are also incorporated to provide textual or verbal support to diabetes and heart patients. The patients can interact with tele-facilitators and chatbot to clear their doubts in case of emergency or requirements. The granular computing assists in improving the predictive accuracy of data mining procedures in the early diagnosis of these diseases. The granular computing tries to find effective data granules so that the system can be precisely trained to make better predictions. The sentiment analyzer is used to mine social media and different online sources to determine better and updated information for diabetes and cardio patients. Sentiment analyser is able to perform discourse and pragmatic analysis so that the complex and compound sentence can also be effectively mined. Figure 2 represents the 3-Tier working model of SMDIoT. Tier 1 and 3 are related to patients (with embedded bio-sensors) and healthcare professional respectively. The middle tier is the heart of SMDIoT. Middle tier contains two types of components namely Data, IoT, Telehealth facilitator, chatbots, context entity search, sentiment analyzer, granular computing, diagnosis procedure, data analysis and visualization. The innovation of SMDIoT lies in the assimilation of granular computing, chatbots, intelligent search and sentiment analyzer.

| Features of SDMIoT |
|---|
| It is a completely connected network of physical and cyber world objects that are supposed to give 24X7X365 connectivity. |
| - Patients data will be regularly collected and monitored. |
| - The data will be regularly transmitted to both patients and healthcare professionals. |
| - In case of not availability of healthcare professional, the patient will be able to interact with a chatbot to solve his/her query. |
| - Context entity search module will expedite and improves the online search process of the patients. |
| - Data mining procedures will be used to early disease diagnosis and other critical issues. |
| - Sentiment analysis will assist in precise mining diabetes and heart-related information from the social network sites. The same information will be transmitted to both patients and healthcare professional for their reference. |
| - Granular computing will assist to work on granules of the data so that the predictive rate of early disease diagnosis can be further improved. |
| - The use of sensor will provide the facility for remote monitoring of the patient if required. |

## 4.1 Components of SDMIoT

For real implementation of the system, firstly, all different components of the cyber and physical world should be interconnected using IoT. For easy tracking, all cyber and physical entities of devices should possess a unique identification. After connecting cyber and physical entities, the role of DM and machine learning (ML) comes into play. DM and ML are liable for designing intelligent procedure for early diagnosis, chatbot design, sensor training, auto call training etc. Finally, meticulousness and performance of the system are enhanced by incorporating the features of chatbots, granular computing, sentiment analysis and contextual entity search. The need and working of different components of SMDIoT are given below:

- Patient: Both diabetes and cardiovascular patients are the major components of SMDIoT. The patients can exchange their health-related data with this smart and remote monitoring system through different sensors and other medical devices.

- A Bio-sensor: deployed into the body of the diabetic or cardio patient so that the regular and automatic health monitoring of the body can take place. Based upon fluctuation, an automatic reminder will be sent to both patients as well as a concerned healthcare profession. SMDIoT will provide complete activity log so that the patient can also scrutinize his/her health status. Furthermore, by implementing wearable bio-sensors, SMDIoT can also detect metabolic activities of the patient and can instruct the patient to change his/her routine to make him/her safe.

- Internet: Internet is the key requirement for SMDIoT as all physical and cyber world objects are connected through internet only.

- Social Network Sites: For semantic analysis, the data related to diabetes and cardio diseases is collected from different social network sites like Facebook, Twitter, WhatsApp, Instagram etc.

- Tele Facilitator: In case of emergency, SMDIoT is proficient in making an auto call to ambulance service. For better psychological and psychosocial conditions, the system is also competent to provide telehealth facility, i.e. counselling through telephone or mobile.

- Insulin Detector: This component is responsible for detecting the need for insulin and to transmit the notification to the patient. At the advanced level, smart wearable devices can be linked with the system that will automatically inject the required level of insulin.

- Cardiovascular Sensor (CS): it is supposed to record BP, respiration rate, weight and the other status of the heart. CS is capable of transmitting vital information of the cardio patient to SMDIoT. The data extracted using CS will then be processed, and again resultant will be transmitted to both concerned patient and healthcare professionals.

- Data: The complete and regular data related to the different condition of both diabetes and heart patients are recorded. Data is of utmost importance as it represents the safe or critical state of the patients.

- Chatbot: Regrettably, it is difficult to consult with a doctor, mainly if one requires counselling on non-life intimidating problems. In SMDIoT, the chatbot facility is also supported so that in case of non-availability of the healthcare experts, one can interact with expert machines for assistance. A chatbot is a machine that poses some queries to the patient simply by reshuffling what the patient himself alleged.

- Context Entity Search: Context entity search is a confluence of intelligence, flexibility and performance. With content entity search, one is permitted to mine content from structured as well as unstructured data effectively. With this one can more focus on the required material as it provides the facility of scope search, where search can be focused on a paragraph or even to specific sentences rather than the whole document. Sentences and paragraph are treated as a granule rather than the whole document or file. With context entity search module, one can

- Sentiment Analyzer: This module is incorporated in SMDIoT to provide the facility of sentiment analysis related to diabetes and cardio disorders. With this feature, the user can effectively mine diabetes and cardio related information from a momentous volume of information available over social media and internet. The information can be stocked up and used for training purpose for diagnosis procedure. SMDIoT can-do pragmatic as well as discourse analysis. However, the multilingual features have not been incorporated yet. The extracted and precise information will then be automatically transmitted to both patients and healthcare professionals.

- Granular Computing: Granular computing works with granules. Granules are a cluster of analogous objects. These are clustered together due to the features like similarity, functionality, location, coherence etc. Granular computing assists in achieving more precise results as it works on the representative of data rather than data itself. Therefore, for developing smart and advanced healthcare system (SMDIoT), granular computing is used. Two different pyramids of granular information related to diabetes and cardio diseases are constructed.

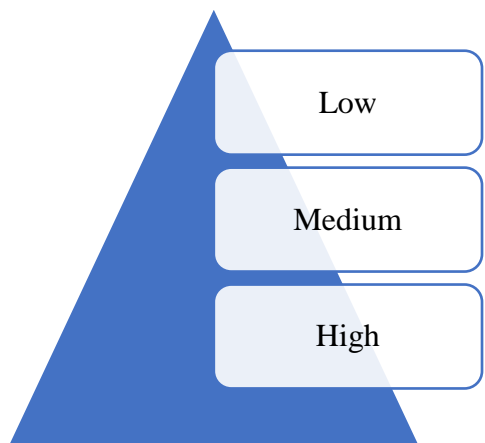

**Diabetes Granule Pyramid**

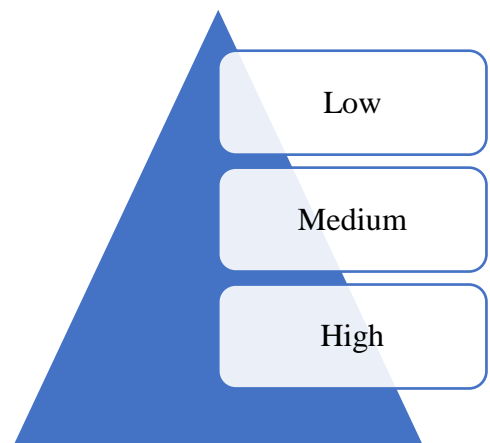

**Cardio Granule Pyramid**

These pyramids arrange different information granules related to diabetes and cardio disorders. Pyramid may contain some indexed layers based upon size and dimension of information granules. The use of different sized information granules helps in effective training of disease diagnose procedure. Therefore, the use of these granular pyramids in SMDIoT will improve the predictive accuracy and the overall performance of SMDIoT. It will also assist in exchanging the information between patients, healthcare professional and system itself.

- Early Diagnosis procedure: A precise and intelligent hybrid procedure for the early diagnosis of diabetes and cardio disorder is designed. The principle of the developed procedure is based upon Naive Bayes, information theory and genetic algorithm (GA). Here, information theory is used to remove uncertainty and GA will expedite the diagnosis process.

- Data Analysis: SMDIoT is trained enough to analyze diabetes and heart patient's data. The system is able to classify the state of the patient. The complete data analysis of the patients is transmitted to both patients and healthcare professional for their reference and further action.

- Data Visualization: It seems to be unrealistic to ask a healthcare professional to pore over the capacious data or directly examined the data received from the IoT-based biosensors embedded in the human beings. Rather, for effective clinical practice, the results from the proposed system need to be depicted to the physicians in the perceptive format so that it becomes readily comprehended. Therefore, in the proposed system several visualization techniques have been employed so that the data should be normalized before presenting it to the physicians.

SDMIoT seems to be more effective and smart solution for complete diabetes care. The framework is supposed to provide 24X7X365 connectivity between patients, healthcare professionals and diabetes monitoring devices. The novelty of SDMIoT lies in combining different computational procedures viz., machine learning, IoT and smart technologies (granular computing, chatbot, contextual entity search, sentiment analysis). In addition, all economic losses related to hospital visits, standing in a queue, travelling costs, leave from office etc. can be minimized using SDMIoT.

## 4.2 Challenges in SMDIoT

One of the most critical issues for SMDIoT is the security and privacy of patient's data. One has to make data secure from unauthorized and unauthenticated attacks. In SMDIoT, most of the devices are battery operated and based upon Wi-Fi connectivity. Therefore, battery consumption will be another issue. Additionally, mobile devices will create a signal problem due to a lot of metal and walls available in hospital or clinics.

## 5. Conclusion

Diabetes and cardio problems are two critical human disorders. A significant number of people died due to these critical human disorders. As per statistics, cardio patients have a maximum rate of death. The past research reveals that different data mining techniques such as Naive Bayes, random forest, C4.5, support vector machine, neural network etc. have been extensively used in the diagnosis of several human disorders. A good rate of classification has been achieved for both diabetic and cardiovascular patients. Additionally, some of the authors have also devised IoT-based healthcare systems. However, to compete with current technology and to save the life of patients, a more smart and intelligent system is needed. Therefore, a hybrid artificial intelligent and smart framework based upon Data Mining, IoT, chatbots, contextual entity search, sentiment analysis and granular computing has been proposed. Therefore, here, a smart data mining and IoT (SMDIoT) based advanced healthcare system for proficient diabetes and cardiovascular diseases have been proposed. The hybridization of data mining and IoT techniques are supposed to give an effective and economical solution to diabetes and cardio patients. With the use of SMDIoT, vital medical resources and manpower will be better utilized; hence, the patient care will be further enhanced. In spite of healthcare, it will substantially reduce economic losses of diabetes and cardio patients by minimizing various implicit and explicit medical expenses.

In future work, the proposed system will be implemented. Additionally, other emerging computing techniques like Ant Lion Optimization, Whale optimization algorithm, Moth-flame optimization, Crow Search Optimization may also be used to improve the predictive rate of diagnosis. In spite of this, more research work is needed to generate smart and secure healthcare diagnostic systems. Moreover, as stated earlier, most of the devices will be battery operated. Therefore, battery consumption should be optimized.